\documentclass[12pt]{article}
\usepackage{amsmath}
\usepackage{amssymb}
\usepackage{amscd}
\usepackage{amsfonts}
\usepackage{amsthm}
\usepackage{authblk}

\usepackage{mathrsfs}
\usepackage[matrix,arrow,curve]{xy}
\usepackage{amsbsy}
\usepackage{slashed}
\usepackage{amssymb}
\usepackage{bbm}
\makeindex
  \newtheorem{theorem}{Theorem}[section]
  \newtheorem{df}[theorem]{Definition}

  \newtheorem{lemma}[theorem]{Lemma}
  
\theoremstyle{remark}
  
  \newtheorem{rem}[theorem]{Remark}

\def\bea{\begin{eqnarray}}
\def\eea{\end{eqnarray}}

 \newcounter{biscompt}

 \theoremstyle{plain}

\usepackage{hyperref}

 \numberwithin{equation}{section}

\newcommand{\Dj}{\hbox to 8pt{\raisebox{.4\height}{-}\hss D}}

\newcommand{\tc}{\ensuremath{\mathscr T}}

\newcommand{\beq}[1]{\begin{equation}\label{#1}}
\newcommand{\eq}{\end{equation}}
\newcommand\eeq{\end{equation}}
\newcommand\bqa {\begin{eqnarray}}
\newcommand\eqa {\end{eqnarray}}

\newcommand\nn {\nonumber}

\newcommand{\bear}{\begin{array}}
\newcommand{\enar}{\end{array}}

\newcommand{\R}{\mathbb{R}}

\newcommand{\slnr}{\ensuremath{\mathfrak{sl}_n(\mathbb R)}}
\newcommand{\glnr}{\ensuremath{\mathfrak{gl}_n(\mathbb R)}}
\newcommand{\sonr}{\ensuremath{\mathfrak{so}_n(\mathbb R)}}

\newcommand{\bp}{\ensuremath{\mathfrak{b}^+_n(\mathbb R)}}
\newcommand{\bm}{\ensuremath{\mathfrak{b}^-_n(\mathbb R)}}

\newcommand{\Slnr}{\ensuremath{SL_n(\mathbb R)}}
\newcommand{\Sonr}{\ensuremath{SO_n(\mathbb R)}}
\newcommand{\Bp}{\ensuremath{B^+_n(\mathbb R)}}
\newcommand{\Bm}{\ensuremath{B^-_n(\mathbb R)}}
\newcommand{\sy}{\ensuremath{Symm_n(\mathbb R)}}

\newcommand{\tr}{\mathrm{tr}\,}

\usepackage[usenames]{color}

\usepackage{tikz}

\usepackage{graphicx}

\title{ The Lie\,-Bianchi integrability of the full symmetric Toda system}
\author[1,2,3]{Yury B. Chernyakov\footnote{chernyakov@itep.ru}}
\author[1,2,4]{Georgy I. Sharygin\footnote{sharygin@itep.ru}}
\author[1,4,5]{Dmitry V. Talalaev\footnote{dtalalaev@yandex.ru}}
\affil[1]{\small NRC ``Kurchatov institute'', Kurchatov square, 1, 123182, Moscow, Russia}
\affil[2]{\small  MITP, Institutskii per. 9, 141700, Dolgoprudny, Russia}
\affil[3]{\small 
IITP RAS, Bolshoy Karetny per. 19, build.1, 127051, Moscow, Russia}
\affil[4]{\small  Lomonosov MSU, Leninskie Gory, 1, 119991, Moscow, Russia}
\affil[5]{\small Demidov YarSU, Sovetskaya Str. 14, 150003, Yaroslavl, Russia}
\date{}

\begin{document}

\maketitle

\begin{abstract}
In this paper we prove that the full symmetric Toda system is integrable in the sense of the Lie-Bianchi criterion, i.e. that there exists a solvable Lie algebra of vector fields of dimension $N=\dim M$ on the phase space $M$ of this system such that the system is invariant with respect to the action of these fields. The proof is based on the use of symmetries of the full symmetric system, which we described earlier in \cite{CSS23}, and the appearance of the structure of the stochastic Lie algebra in their description.
\end{abstract}
\section{Introduction}
Recall that \textit{full symmetric Toda system} is determined by the following Lax equation: let $L=L(t) \in \sy$ be a real symmetric matrix, which we can without loss of generality assume to have zero trace\footnote{One can show that the trace is in effect Casimir function of the corresponding Poisson structure.}. Then we define its na\"\i ve antisymmetrization $M(L)$ as the difference $L_+-L_-$, where $L_\pm$ are the upper/lower triangular parts of $L$. In these terms the full symmetric Toda system is given by the equation
\[
\dot L=[M(L),L].
\]
This differential equation is quite remarkable due to a long list of wonderful properties that it possesses. First of all this equation is in fact Hamilton equation with respect to the Poisson structure on \sy, induced from its identification with $\bm^*$, corresponding to Hamilton function $H=\frac12\tr(L^2)$ (see \cite{Ar1979}-\cite{S1980}). This system is integrable in the sense of Liouville's criterion, i.e. there exists a commutative family of first integrals of this system, whose functional dimension is greater or equal to one half of the dimension of phase space, \cite{DLNT}. Moreover this system is super-integrable in the sense of Nekhoroshev, \cite{Neh1972} (see \cite{GS1999}-\cite{CS15}); in effect every \Bp-invariant function on \slnr\ induces an integral of this system, when restricted to \sy, \cite{EFS1993},\,\cite{CS25}. Further, solutions of this equation can be obtained by the QR-decomposition of the exponent of initial position, \cite{TCS23}, its singular points are enumerated by permutations of $n$ elements and its trajectories connect these points. Moreover, the trajectories connect these points and go in accordance with the Bruhat order on permutations: from singular points corresponding to permutations that stand lower in this order to those that stand higher, \cite{CSS14}. One can further generalise this system to obtain systems on ``generalized symmetric matrices'', i.e. on subspaces in real semisimple Lie algebras, orthogonal to maximal compact subalgebras in them; in this case the trajectories will connect points indexed by the elements of Weil group and go in accordance with the Bruhat order in that group, from lower to higher points, \cite{CSST20}.

In this paper we are going to prove yet another remarkable property of the full symmetric Toda system. Recall that in the theory of dynamical systems there exist few integrability criteria, where ``integrability'' is to be understood as the possibility to express the solutions of a system of differential equation by formulas that contain compositions of the operations of taking integrals (passing to ``quadratures'' in the terminology of the XIX-th century) and applying elementary functions (such as polynomials, exponents, trigonometric functions etc.) to the coefficients of the initial equations. Liouville's theorem, mentioned above is one of such criteria, there also exists its generalization by Nekhoroshev (also mentioned above), where the number of first integrals should be greater, but the commutativity condition is relaxed. Both these integrability criteria are based on the use of first integrals of the system.

An alternative approach to integrability, going back to the works of Sophus Lie, emerges from the considerations of infinitesimal symmetries of the system, i.e. from the properties of the Lie algebra of (local) vector fields on the phase space of the system, commuting with the system. It turns out that this point of view can also yield a criterion for the integrability ``in quadratures'', the \textit{Lie-Bianchi criterion}:
\begin{theorem}[Lie-Bianchi theorem]\label{theo:lie-bian}
If a dynamical system on $\mathbb R^n$ has a solvable Lie algebra of infinitesimal symmetries, whose dimension is $n-1$ and whose action on the space is locally free, then the system is integrable in quadratures.
\end{theorem}
Here local freeness means that the basis fields $Z_1,\dots,Z_{n-1}$ in this Lie algebra together with the field $X$ generate $T_{y'}\R^n$ at all points $y'$ in a vicinity of $y$. For the proof of this theorem we refer to the book \cite{Bir1963} (see also the survey article \cite{Koz1983}, where it is treated in a more modern way). Clearly, the same statement holds for dynamical systems on arbitrary manifolds. Also observe that the Lie algebra, spanned by these fields needs not have structure constant independent on the points, although it is hard to imagine finite-dimensional Lie algebra with this property.

The undoubted advantage of this approach is that the conclusion of Lie-Bianchi theorem is not restricted by the assumption that the system in question is Hamiltonian; in effect this theorem can be applied to a large variety of systems, which undoubtedly gives one a great freedom of action. On the other hand, the conditions on vector fields that one should verify are quite restrictive, and the number of fields that one has to find should be almost equal to the dimension of the manifold, which is twice the number of first integrals that one has to find in Liouville's criterion (see above). That is probably the reason, why this statement is much more rarely used in practice than Liouville's.

It turns out (as we will prove it in this paper) that the full symmetric Toda system does satisfy the criterion of Lie-Bianchi theorem, i.e. it has a solvable algebra of symmetries of dimension equal to the dimension of the phase space, the space of traceless real symmetric matrices, which is $\frac{n(n+1)}{2}-1$. Moreover, it has a finite-dimensional Lie algebra of infinitesimal symmetries (given by vector fields with rational coefficients), isomorphic to a central extension of the Lie algebra of \textit{stochastic matrices}. In effect, it has many other infinitesimal symmetries, which form finite-dimensional Lie algebras; these symmetries were constructed earlier from representations of the Lie algebra $\mathfrak{sl}_n$ by the first and the second author in collaboration with A.Sorin, (see \cite{CSS23}, however there we didn't check that these symmetries are closed under commutators). The relation of the Lie algebra of symmetries with the stochastic Lie algebra, which we establish here, allows one to find solvable subalgebras that ensure Lie-Bianchi integrability in a rather simple way.

The composition of this paper is as follows: we begin by recalling in section \ref{sect:vecfiel1} the main steps of the construction from \cite{CSS23}, which gives a large collection of vector fields, commuting with Toda system. As we have mentioned, these fields are closely related to the representations of $\mathfrak{sl}_n$. We then prove by a direct computation (see section \ref{sect:vecfiel2}) that the symmetries, corresponding to the tautological representation of $\mathfrak{sl}_n$ are in fact closed under commutation. This method can in effect be used for many other representations , e.g. for the representation in exterior square of $\mathbb R^n$, thus yielding a series of finite-dimensional Lie algebras of symmetries for the Toda system, see the appendix \ref{higheralg}. However, these algebras are quite hard to study, so we restrict our attention on the first one. We show that this algebra does in fact contain a solvable subalgebra, which acts locally-free at a generic point in $SO_n(\R)$, this is done by observing that the Lie algebra in question is isomorphic to the central extension of the subalgebra of ``stochastic'' matrices in $\mathfrak{gl}_n$, see section \ref{stochastic}.

\section{Infinitesimal symmetries of the Toda system}\label{sect:vecfiel1}
Let us recall the construction of infinitesimal symmetries for the full symmetric Toda system from \cite{CSS23}. Here we will restrict our attention to the case of system on the space of usual real symmetric matrices, although in the paper \cite{CSS23} this construction was given for the analogous system defined for arbitrary Cartan pair decomposition
\[
\mathfrak g=\mathfrak k\oplus\mathfrak p
\]
of a semisimple real Lie algebra $\mathfrak g$. Here $\mathfrak k$ is a maximal compact subalgebra in $\mathfrak g$ and $\mathfrak p$ is its orthogonal complement with respect to the Killing form (instead of Killing form one can take some other nondegenerate invariant pairing on $\mathfrak g$, which is positive/negative definite on $\mathfrak k$).

So we let $\mathfrak g=\mathfrak{sl}_n(\R)$, the space of all traceless real matrices of size $n\times n$. By a slight abuse of definitions we will consider the nondegenerate invariant form on $\mathfrak{sl}_n(\R)$ given by
\[
\langle A, B\rangle=\mathrm{tr}(A^tB),\ A,B\in\mathfrak{sl}_n.
\]
It differs from the usual Killing form on $\mathfrak{sl}_n(\R)$ by a constant factor and the transposition of the first argument, which will not influence the resulting formulas in any  essential way. If now $\mathfrak k=\mathfrak{so}_n(\R)$ is the space of real skew-symmetric $n\times n$ matrices then we see that its orthogonal complement $\mathfrak p=\mathfrak{so}_n^\perp$ with respect to this form is equal to the space of traceless symmetric matrices $Symm_n(\mathbb R)$: indeed the pairing $\langle A,B\rangle$ vanishes for all pairs of symmetric matrix $A$ and antisymmetric matrix $B$, so the conclusion follows by dimension counting.

\subsection{Vector fields $\tc^X$ on \Sonr}
We are going to recall the construction of infinitesimal symmetries of the Toda system from the paper \cite{CSS23} in this case; for some details of the proof as well as for the form of this construction in general case the interested reader is referred to the cited text.

We begin by observing that the action of $SO_n(\mathbb R)$ on the space of symmetric matrices by conjugation allows one to transfer the vector fields on the former Lie group to vector fields on symmetric matrices: indeed every matrix $L\in Symm_n(\mathbb R)$ can be represented as $\Psi\Lambda\Psi^{-1}$ where $\Lambda$ is real diagonal and $\Psi\in SO_n(\mathbb R)$. Moreover, this representation for a generic $L$ is unique up to a permutation of eigenvalues in $\Lambda$ and multiplication of $\Psi$ by a diagonal matrix with eigenvalues $\pm1$. Thus this representation can be chosen in a unique way locally, giving a local finite leaf covering of $Symm_n(\mathbb R)$ by $SO_n(\mathbb R)\times\mathbb R^{n-1}$:
\[
(\Psi,\Lambda)\mapsto\Psi\Lambda\Psi^{-1}
\]
and conclusion follows. A more instrumental way to view this construction is as follows: if $\Psi(t)$ is a trajectory of the vector field $\xi$ on $SO_n(\R)$, then its image on $Symm_n(\R)$ will have trajectories $\Psi(t)\Lambda\Psi(t)^{-1}$.

So it is sufficient for our purposes to find a suitable collection of vector fields on the group $SO_n(\mathbb R)$. To this end let $X\in\mathfrak{sl}_n$ be an arbitrary matrix. We will associate with $X$ a smooth vector field $\mathscr T^X$ on $SO_n(\mathbb R)$: for a point $\Psi\in SO_n(\mathbb R)$ we identify $T_\Psi SO_n(\mathbb R)$ with right translations by $\Psi$ of anti-symmetric matrices and put
\[
\mathscr T^X(\Psi)=M(\Psi X\Psi^{-1})\Psi,
\]
where $M$ is the natural projection $\mathfrak{sl}_n(\R)\to\mathfrak{so}_n(\R)$ in the direct sum decomposition
\begin{equation}\label{eq:decompgau}
\mathfrak{sl}_n(\R)=\mathfrak{so}_n(\R)\oplus\mathfrak b_n^-(\R),
\end{equation}
where $\mathfrak b_n^\pm(\R)$ are the Borel subalgebras of upper/lower triangular matrices in $\mathfrak{sl}_n(\R)$.

Before we proceed, remark that the map $M:L\mapsto M(L)=L_+-L_-$ from the introduction is equal to the restriction of projection $M$ onto the space of symmetric matrices. Moreover, let $\Lambda$ be a diagonal matrix, then the field $\mathscr T^\Lambda$ will induce Toda system on $Symm_n(\R)$: indeed, differentiate the corresponding curve $L(t)=\Psi(t)\Lambda\Psi(t)^{-1}$ on $Symm_n(\R)$ (where $\Psi(t)$ is a trajectory of $\mathscr T^\Lambda$) then $\dot\Psi=\mathscr T^\Lambda(\Psi)=M(L)\Psi$ and since
\[
0=\frac{d}{dt}(1)=\frac{d}{dt}(\Psi\Psi^{-1})=\dot\Psi\Psi^{-1}+\Psi\dot\Psi^{-1}
\]
we get $\dot\Psi^{-1}=-\Psi^{-1}M(L)$, so
\[
\begin{split}
\dot L&=\dot\Psi\Lambda\Psi^{-1}+\Psi\Lambda\dot\Psi^{-1}\\
         &=M(L)\Psi\Lambda\Psi^{-1}-\Psi\Lambda\Psi^{-1}M(L)=[M(L),L].
\end{split}
\]
Hence the field $\mathscr T^\Lambda$ induces Toda system on the space of matrices with fixed spectrum. On the other hand the evolution of Toda system clearly preserves the spectrum of matrix $L$, thus we conclude that the trajectories of Toda system on the space of matrices whose spectrum is given by $\Lambda$ coincide with the trajectories of the fields $\mathscr T^\Lambda$ on the same space. For this reason we will call $\mathscr T^\Lambda$ the \textit{Toda field}. Now as we saw earlier, symmetries of the Toda system on $Symm_n(\R)$ can be obtained from the vector fields on $SO_n(\R)$, commuting with Toda field.

Let us say a few words about the properties of fields $\mathscr T^X$ for arbitrary $X$. Since both terms of decomposition \eqref{eq:decompgau} are subalgebras in $\mathfrak{sl}_n(\R)$, the projection $M$ satisfies the equality
\begin{equation}\label{eq:Nij}
M^2([X,Y])-M([M(X),Y])-M([X,M(Y)])+[M(X),M(Y)]=0.
\end{equation}
This formula allows one to calculate the commutator $[\mathscr T^X,\mathscr T^Y]$: we use the remark that the commutator of fields $\xi(\Psi)=x(\Psi)\Psi,\,\eta(\Psi)=y(\Psi)\Psi,\, x(\Psi),y(\Psi)\in\mathfrak{so}_n(\R)$ on $SO_n(\R)$ is equal to $z(\Psi)\Psi$, where
\[
z(\Psi)=\xi(y)(\Psi)-\eta(x)(\Psi)-[x(\Psi),y(\Psi)].
\]
Then, applying this formula to $\mathscr T^X,\mathscr T^Y$ we have $x(\Psi)=M(\Psi X\Psi^{-1}),y(\Psi)=M(\Psi Y\Psi^{-1})$ so, using equation \eqref{eq:Nij}, we get
\[
\begin{split}
z(\Psi)&=M([M(\Psi X\Psi^{-1}),\Psi Y\Psi^{-1}]-M([M(\Psi Y\Psi^{-1}),\Psi X\Psi^{-1}]\\
          &\quad-[M(\Psi X\Psi^{-1}),M(\Psi Y\Psi^{-1})]=M^2(\Psi[X,Y]\Psi^{-1}).
\end{split}
\]
Since $M^2=M$ (in fact $M$ is a projector), we have
\[
[\mathscr T^X,\mathscr T^Y]=\mathscr T^{[X,Y]}.
\]
Thus, we conclude that \textit{vector fields $\tc^X,\,X\in\slnr$ give a representation of the Lie algebra \slnr\ on the space \Sonr}.

\subsection{Vector fields $T^{ij}$}\label{sect:vecfiel3}
Consider the fields $\tc^X$ and $\tc^Y$ for $X=\Lambda$, and $Y=E_{ij}$, the matrix unit (here $i$ is the number of row and $j$ the number of column where $1$ stands); we get
\[
[\mathscr T^\Lambda,\mathscr T^{E_{ij}}]=\mathscr T^{[\Lambda,E_{ij}]}=(\lambda_i-\lambda_j)\mathscr T^{E_{ij}}.
\]
So in order to obtain the symmetries of Toda systems we have to multiply the fields $\mathscr T^{E_{ij}}$ by suitable rational functions $F_{ij}$ of $\Psi$, for which
\[
\mathscr T^\Lambda(F_{ij})=(\lambda_j-\lambda_i)F_{ij}.
\]
 This can be done by virtue of the following observation: suppose that $\rho:\slnr\to\mathrm{End}(V)$ is a real representation of \slnr. We will use the same letter to denote the representations of the algebras \sonr,\,\bp,\,\bm\ and the groups \Slnr,\,\Sonr\ and \Bp,\,\Bm, induced from $\rho$. Without loss of generality we can assume that $\rho$ is a ``minimal weight'' orthogonal representation, i.e. that there exists a Euclidean scalar product $\langle,\rangle$ on $V$, invariant with respect to the action of $\rho(\Sonr)$ and we can choose an orthonormal basis $v_1,v_2,\dots,v_N=v_-$ of $V$, for which the action of Cartan subalgebra $\mathfrak h_n$ (the subalgebra of diagonal matrices in \slnr) is diagonal, i.e.
 \[
 \rho(\Lambda)(v_k)=\omega_k(\Lambda)v_k,\,k=1,\dots,N
 \]
 for some weights $\omega_k\in\mathfrak h_n^\ast$ and
 \[
 \rho(X)(v_-)=0,\ \mbox{for all}\ X\in\mathfrak n_-,
 \]
where $\mathfrak n_-$ is the nilpotent subalgebra of strictly lower-traingular matrices in \slnr. Then we put
 \begin{equation}\label{eq:functionsi}
 f_i^{\rho}(\Psi)=\langle \rho(\Psi)(v_i),v_-\rangle,\ \Psi\in\Sonr,\ i=1,\dots,N.
 \end{equation}
Then a straightforward computation in \cite{CSS23}, using the properties of $\rho$ shows that
 \[
 \mathscr T^\Lambda (f_i^{\rho})(\Psi)=(-\omega_i(\Lambda)+g^\rho_\Lambda(\Psi))f_i^\rho(\Psi),
 \]
for some smooth function $g^\rho_\Lambda$ on \Sonr. The important fact is that this function depends on $\rho$ and $\Lambda$ but not on the index $i$! This suggests the following idea: put $F^\rho_{ij}=\frac{f_i^\rho}{f_j^\rho}$ then
 \[
  \mathscr T^\Lambda (F_{ij}^{\rho})(\Psi)=(\omega_j(\Lambda)-\omega_i(\Lambda))F_{ij}^\rho(\Psi).
 \]
If it happens that $\omega_j(\Lambda)-\omega_i(\Lambda)=\lambda_j-\lambda_i$, then the product $F_{ij}^{\rho}\mathscr T^{E_{ij}}$ will commute with Toda field:
\[
[\mathscr T^\lambda,F_{ij}^{\rho}\mathscr T^{E_{ij}}]=0.
\]
One particular choice of the data for this construction would be $V=\R^n$ with tautological representations and equipped with the standard Euclidean structure, so that the canonical basis is orthonormal. Then $f_i^\rho(\Psi)=\psi_{ni}$ and $F_{ij}^\rho=\frac{\psi_{ni}}{\psi_{nj}}$. We will denote the corresponding vector fields $F_{ij}^{\rho}\mathscr T^{E_{ij}}$ simply by $T^{ij}$ in this case.
\begin{rem}\label{rem:rem1}
There is a geometric point of view on the construction of fields $\mathscr T^{X},\ X\in\slnr$: consider the principal \Sonr-bundle $\Slnr\to\Sonr\backslash\Slnr$ with contractible base space $\Sonr\backslash\Slnr$. In effect, one knows that the quotient $K\backslash G$, where $K\subset G$ is a maximal compact subgroup in $G$ and $G$ is semisimple is always diffeomorphic to a Euclidean space, see Helgason's book \cite{Helg1978}. In our case one can regard the quotient $K\backslash G= \Sonr\backslash\Slnr$ as the space of real symmetric positive-definite matrices with unit determinant $SSymm_n^+(\R)$. In the terms of matrices, this principal bundle is equivalent to the polar decomposition:
\[
\Slnr\cong\Sonr SSymm_n^+(\R),\ X=QR,\ X\in\Slnr,\,Q\in\Sonr,\,R\in Symm_n^+(\R)
\]
Observe that every matrix in $SSymm_n^+(\R)$ is equal to the exponent of a traceless matrix in \sy.
So $SSymm_n^+(\R)$ is a Euclidean space and hence is contractible. It follows that the bundle $\Slnr\to SSymm^+_n(\R)$ can be trivialized (this follows from the contractibility of the base $SSymm_n^+(\R)$, but can be proved independently). So via the exponent map we get a projection
\[
\xymatrix{
{\Slnr\cong\Sonr\times Symm_n^0(\R)}\ar[d]_{\Sonr}\\
Symm_n^0(\R).
}
\]
Recall that a principal connection on a principal $G$-bundle $P\to B$ (with left action of $G$) is determined by a connection $1$-form, the $\mathfrak g$-valued differential $1$-form $\omega$ on $P$ that satisfies the following two conditions:
\begin{enumerate}
\item[(\textit{i})] Invariance: $L_g^*(\omega)=Ad_g(\omega)$ for all $g\in G$, where $R_g$ denotes the action of $g$ on $P$ (here $L_g$ is the left translation by $g\in K$ on $P$);
\item[(\textit{ii})] Normalization: $\omega(\tilde X)=X$ for all $X\in\mathfrak g$, where $\tilde X$ denotes the vertical vector field on $P$, induced by $X$:
\[
\tilde X(p)=\frac{d}{dt}_{|_{t=0}}(\exp(tX)\cdot p),\ p\in P.
\]
\end{enumerate}
If we omit the first condition, and modify the definition so that $\omega$ will be regarded as a $1$-form on $TP$ with values in \textit{vertical subbundle} $T^vP$ of $P$ (one can define $T^vP$ it as the subbundle of $TP$, spanned by the vectors, tangent to fibres of the bundle). In this way we will get the notion of a generic \textit{Ehresmann's connection}, see \cite{Michor}. In this approach the curvature of connection is determined by the \textit{Fr\"olicher-Nijenhuis bracket} $R=\frac12[\omega,\omega]$.

We can now use the projector $M:\slnr\to\sonr$ to induce such a form: we put
\[
\omega(\tilde\xi(g))=\widetilde{M(\xi)}(g),\,g\in\Slnr,\,\xi\in\slnr.
\]
Here $\tilde\xi$ is the right-invariant vector field on \Slnr, induced by a vector $\xi\in\slnr$; in matrix terms $\tilde\xi(g)=\xi\cdot g$; we extend it to whole $T\Slnr$ in a unique way, since right-invariant fields give a global basis in $T\Slnr$.

Clearly, the map $\omega$ is well defined, takes values in the tangent bundle, induced by (left) \Sonr-orbits in \Slnr\ and for every $\eta\in\sonr$ we have
\[
\omega(\tilde\eta(g))=\tilde\eta(g),\ g\in\Slnr.
\]
Hence $\omega$ is a connection. Also observe that for every left-invariant vector field $\hat\xi(g)=g\cdot\xi,\,\xi\in\slnr,\,g\in\Slnr$ we have
\[
\omega(\hat\xi(g))=\widetilde{M(Ad_g(\xi))}(g).
\]
In case $g\in\Sonr\subset\Slnr$ we get $\omega(\hat\xi(g))=\mathscr T^\xi(g)$.

Also observe that the connection $\omega$ is flat: we compute by the definition of Fr\"olicher-Nijenhuis bracket
\[
\begin{split}
[\omega,\omega](\tilde X,\tilde Y)&=\omega([\tilde X,\tilde Y])-\omega([\omega(\tilde X),\tilde Y])\\
&\quad-\omega([\tilde X,\omega(\tilde Y)])+[\omega(\tilde X),\omega(\tilde Y)]=0
\end{split}
\]
for all right-invariant fields $\tilde X,\tilde Y$ due to the same equation for $M$.
\end{rem}

\section{The Lie algebra of symmetries of the Toda system}\label{sect:vecfiel2}

Recall, that the $T^{ij}$ denotes the vector field on \Sonr\ given by the formula
\[
T^{ij}= \frac{\psi_{ni}}{\psi_{nj}}\tc^{E_{ij}},
\]
see the discussion preceding the remark \ref{rem:rem1}. As we explained in section \ref{sect:vecfiel3}, this field commutes with the Toda fields $\tc^\Lambda$ on the special orthogonal group and hence induces symmetries of the full symmetric Toda system. In terms of the standard coordinates on the group \Sonr\ we can compute explicitly: let $\Psi=(\psi_{ij})$ be an orthogonal matrix, then
\beq{Toda-action}
\tc^{\Lambda} \psi_{ni} = (a_{nn} - \lambda_{i}) \psi_{ni},
\eeq
where $a_{ij}$ are the coefficients of the matrix $L=\Psi\Lambda\Psi^{-1}$, i.e. $a_{nn}=\psi^2_{ni}\lambda_i$.
So the commutator of the fields $\tc^{\Lambda}$ and $T^{ij}$ vanishes:
\[
[ \tc^{\Lambda},T^{ij} ] = \tc^{\Lambda}\left(\frac{\psi_{ni}}{\psi_{nj}}\right)\tc^{E_{ij}} + \frac{\psi_{ni}}{\psi_{nj}}[\tc^{\Lambda},\tc^{E_{ij}}] = (\lambda_{j} - \lambda_{i})\frac{\psi_{ni}}{\psi_{nj}}\tc^{E_{ij}} +  (\lambda_{i} - \lambda_{j})\frac{\psi_{ni}}{\psi_{nj}}\tc^{E_{ij}} = 0.
\]

In order to calculate the commutator of the fields $T^{ij}$ and $T^{kl}$ we must calculate the action of the vector field $\tc^{E_{ij}}$ on functions $f_k^\rho$ for the tautological representation $\rho$ of the algebra \slnr:
\[
f_k^{\rho}(\Psi)=\langle \rho(\Psi)(v_k),v_-\rangle=\psi_{nk}.
\]
We can now use the fact that in tautological representation $\rho(E_{ij})(v_k)=\delta_{jk}v_i$ and further compute 
(see \cite{CSS23} for details):
\beq{Toda-action-2}
\tc^{E_{ij}}\psi_{nk} = \psi_{ni}\psi_{nj}\psi_{nk} - \delta_{ik}\psi_{nj}.
\eeq
and so we get the following formula for the action of the vector field $\tc^{E_{ij}}$ on functions $F_{kl}^\rho=F_{kl}=\frac{\psi_{nk}}{\psi_{nl}}$:
\beq{Function-2}
\tc^{E_{ij}} F_{kl} = - \delta_{ik} F_{jl} + \delta_{il} F_{jl}F_{kl}.
\eeq
For future references it's convenient to set $F_{ii}=1$. Now we can calculate the commutator $[T^{ij} , T^{kl}]$:
\beq{Function-2}
\begin{split}
[T^{ij} , T^{kl}]( \Psi)&=  F_{ij} \tc^{E_{ij}} (F_{kl} \tc^{E_{kl}} ) ( \Psi)- F_{kl} \tc^{E_{kl}} (F_{ij} \tc^{E_{ij}} ) ( \Psi)\\
&=  F_{ij} \tc^{E_{ij}} (F_{kl}) \tc^{E_{kl}}( \Psi) + F_{ij} F_{kl} \tc^{E_{ij}} (\tc^{E_{kl}})( \Psi)\\
&\qquad - F_{kl} \tc^{E_{kl}} (F_{ij}) \tc^{E_{ij}}( \Psi) - F_{kl} F_{ij}  \tc^{E_{kl}} (\tc^{E_{ij}} )( \Psi)\\
&=  F_{ij} (- \delta_{ik} F_{jl} + \delta_{il} F_{jl}F_{kl}) \tc^{E_{kl}} ( \Psi) - F_{kl} (- \delta_{ki} F_{lj}+ \delta_{kj} F_{lj}F_{ij}) \tc^{E_{ij}}( \Psi) \\
&\qquad
+  F_{ij} F_{kl} M (Ad_{\psi}([E_{ij}, E_{kl}]))( \Psi).
\end{split}
\eq
Taking into account, that $[E_{ij}, E_{kl}]=\delta_{kj}E_{il}-\delta_{il}E_{kj}$, that
\[
F_{ij}F_{jl}=\frac{\psi_{ni}}{\psi_{nj}}\frac{\psi_{nj}}{\psi_{nl}}=\frac{\psi_{ni}}{\psi_{nl}}=F_{il}
\]
and so $\delta_{ik}F_{ij}F_{jl}=\delta_{ik}F_{il}=\delta_{ik}F_{kl}$, and similarly
\[
\delta_{il}F_{ij}F_{jl}F_{kl}=\delta_{il}F_{il}F_{kl}=\delta_{il}F_{kl},
\]
since $F_{ii}=1$, we finally get the following formula:
\[
[T^{ij} , T^{kl}] =
- \delta_{ik} F_{kl} \tc^{E_{kl}} + \delta_{il} F_{kl} \tc^{E_{kl}} + \delta_{ki} F_{ij} \tc^{E_{ij}} - \delta_{kj} F_{ij} \tc^{E_{ij}}
+ \delta_{kj} F_{il} \tc^{E_{il}} - \delta_{il} F_{kj} \tc^{E_{kj}}.
\]
Or, in terms of the fields $T^{ij}$ we get
\begin{equation}
\label{comm1}
[T^{ij} , T^{kl}] =
- \delta_{ik} T^{kl} + \delta_{il} T^{kl} + \delta_{ki} T^{ij} - \delta_{kj} T^{ij}
+ \delta_{kj} T^{il} - \delta_{il} T^{kj}.
\end{equation}
A remarkable property of this construction is that \textit{the fields $T^{ij}$ form a finite-dimensional Lie subalgebra in the algebra of rational vector fields on the group \Sonr}.

\subsection{Stochastic Lie algebra and main theorem}
\label{stochastic}
There is an intrinsic relation of the Lie algebra structure (\ref{comm1}) and the so called stochastic Lie algebra. Let us recall its definition.
\begin{df}
The subspace of $n\times n$ matrices  annulling the vector $v_0=(1,1,\ldots,1)\in\R^n$ constitutes a Lie subalgebra $\mathfrak{st}_n(\R)\in \mathfrak{gl}_n(\R)$; this algebra is called \textbf{stochastic} Lie algebra, or the \textbf{algebra of stochastic matrices}.
\end{df}
One can consider the following basis in $\mathfrak{st}_n(\R)$
\bea
f^{ij}=E_{ij}-E_{ii},\qquad \forall i\ne j
\eea
where $E_{ij}$ is the standard $\mathfrak{gl}_n(\R)$ basis (the basis of matrix units). Then
\begin{equation}\label{eq:sth}
\begin{split}
[f^{ij},f^{kl}]&=\delta_{kj}E_{il}-\delta_{il}E_{kj}-\delta_{ki}E_{il}+\delta_{il}E_{ki}-\delta_{kj}E_{ik}+\delta_{ik}E_{kj}\\
&=\delta_{kj}(E_{il}-E_{ii})-\delta_{kj}(E_{ik}-E_{ii})-\delta_{il}(E_{kj}-E_{kk})+\delta_{il}(E_{ki}-E_{kk})\\
&\quad+\delta_{ik}(E_{kj}-E_{kk})-\delta_{ki}(E_{il}-E_{ii})+\delta_{ik}E_{ii}-\delta_{ki}E_{kk}\\
&=\delta_{kj}(E_{il}-E_{ii})-\delta_{kj}(E_{ij}-E_{ii})-\delta_{il}(E_{kj}-E_{kk})+\delta_{il}(E_{kl}-E_{kk})\\
&\quad+\delta_{ki}(E_{ij}-E_{ii})-\delta_{ik}(E_{kl}-E_{kk})\\
&=\delta_{kj}f^{il}-\delta_{kj}f^{ij}-\delta_{il}f^{kj}+\delta_{il}f^{kl}+\delta_{ki}f^{ij}-\delta_{ik}f^{kl}.
\end{split}
\end{equation}
Let us also introduce a Lie algebra $\mathfrak{t}_n$ defined by generators $t_{ij}$ for $i,j=1,\ldots,n$ and relations
\begin{equation}
\label{comm}
[t^{ij} , t^{kl}] =
- \delta_{ik} t^{kl} + \delta_{il} t^{kl} + \delta_{ki} t^{ij} - \delta_{kj} t^{ij}
+ \delta_{kj} t^{il} - \delta_{il} t^{kj}.
\end{equation}
Then
\[
\begin{split}
[t^{ij},t^{kk}]&=- \delta_{ik} t^{kk}+ \delta_{ik} t^{kk}+ \delta_{ki} t^{ij}-\delta_{kj} t^{ij}+\delta_{kj} t^{ik} - \delta_{ik} t^{kj}\\
&= \delta_{ki} t^{kj}-\delta_{kj} t^{ik}+\delta_{kj} t^{ik} - \delta_{ik} t^{kj}=0,
\end{split}
\]
so the linear space $\R^n=\langle t^{ii}\mid i=1,\dots,n\rangle$ is in the center of this Lie algebra. Summing up we get
\begin{lemma}
\label{lemmahom}
The map
\[
\phi: t^{ij}\rightarrow f^{ij}\quad \forall i\ne j;\quad t^{ii}\rightarrow 0  \quad \forall i \nn
\]
is a homomorphism of Lie algebras and moreover we have an isomorphism of Lie algebras
\begin{equation}
\mathfrak{t}_n\simeq \mathfrak{st}_n(\R) \oplus \mathbb{R}^n \hookrightarrow \mathfrak{gl}_n(\R) \oplus \mathbb{R}^n.
\end{equation}
\end{lemma}
\begin{proof}
This statement follows directly from relations (\ref{comm1}) and \eqref{eq:sth} and the fact that the elements $t^{ii}$ are central in $\mathfrak{t}_n$.
\end{proof}

Lemma \ref{lemmahom} provides an obvious way to construct solvable Lie subalgebras in the Lie algebra of vector fields $\langle T^{ij}\rangle$:
\begin{theorem}
If $\mathfrak{g}$ is a solvable Lie subalgebra in $\mathfrak{gl}_n(\R)$ then $\phi^{-1}(\mathfrak{g\oplus \mathbb{R}^n})$ is solvable in $\mathfrak{t}_n.$ Then a canonical homomorphism
\[
\tau:\mathfrak{t}_n\rightarrow \langle T^{ij}\rangle, \qquad \tau:t^{ij}\mapsto T^{ij}
\]
produces a solvable Lie subalgebra of vector fields $\tau\circ \phi^{-1}(\mathfrak{g\oplus \mathbb{R}^n}).$
\end{theorem}
Then we get the following result:
\begin{theorem}
\label{main_theorem}
Let $\mathfrak{g}$ be a solvable Lie subalgebra in $\mathfrak{gl}_n$ (for example, upper Borel subalgebra $\mathfrak{b}_+\subset \mathfrak{gl}_n$) then its preimage $\phi^{-1}(\mathfrak{g}\oplus \mathbb{R}^n)$ in $\mathfrak{st}_n(\R) \oplus \mathbb{R}^n$ provides a solvable Lie subalgebra of symmetries for the Toda system.
\end{theorem}
\begin{rem}
One of conditions of Lie-Binchi criterion is local transitivity of the symmetries, i.e. the condition that the values of corresponding vector fields generate the tangent plane of the phase space of the system at the point. In our case it means that the values of the fields $T^{ij}$ in the image of $\tau$ generate the tangent space of \Sonr. As it is well-known that solvable subalgebras in \glnr\ are all equal to the subalgebras of upper or lower triangular matrices with respect to certain ordering of the basis, we can restrict our attention to the situation to $\mathfrak g=\mathfrak{b}_+$. The image of $\phi^{-1}(\mathfrak{b}_+\oplus \mathbb{R}^n)$ in the Lie algebra of vector fields is spanned by the fields $\langle T^{ij}\mid 1\le i\le j\le n\rangle$. Now it is sufficient to observe that the linear combinations of these fields in a generic point span the same space as the linear combinations of $\mathscr T^{E_{ij}},\,1\le i\le j\le n$. But
\[
\sum_{i,j}\mu_{ij}\mathscr T^{E_{ij}}(\Psi)=M\left(\Psi\left(\sum_{i,j}\mu_{ij}E_{ij}\right)\Psi^{-1}\right)\Psi.
\]
where $\mu_{ij} \in \mathbb{R}$. When $\Psi$ is close to the identity, these fields clearly span the tangent space of \sonr.
\end{rem}

\section{Conclusion}
Let us conclude the text by giving a short ``to do'' list, concerning the constructions we touched on in this paper:
\begin{itemize}
\item
The fields $\mathscr T^X$ were introduced in \cite{CSS23}, where it was shown that suitable coefficients $F_{ij}^\rho$ can be generated from any highest-weight representation. We plan to investigate the corresponding Lie algebras of symmetries in future papers (an example of such algebra is given below in the appendix).
\item
It is well-known (see for example \cite{CSS14}, \cite{CSST20}) that the phase portrait of Toda system reproduces the Hasse diagram of Bruhat order on the permutation group. So it's natural to assume that the structure of the Lie algebra of its symmetries is also closely related to the geometry of the orthogonal group, in particular with the Bruhat stratification of \Sonr. This topic will be at the focus of our future interests.
\item
In \cite{CSS23} we constructed the symmetries of Toda system for arbitrary real form of a semisimple Lie group. So of course, it would be reasonable to ask if the construction of solvable Lie algebra also can be generalized to other root systems.

\item
In addition to the full symmetric Toda system, there's another celebrated variant of ``full Toda'' systems. We mean the well-known \textit{full Kostant-Toda system}. It is well known that the Kostant-Toda system is integrable, moreover, the same version of ``chopping'' integrals (see \cite{DLNT}) is applicable to obtain a complete system of first integrals for that system. We believe that the methods we develop here are also applicable to the  full Kostant-Toda system.

\end{itemize}

\appendix\section{Symmetries of the Toda system induced from representation in $\bigwedge^{2}\R^n$, the case of $n=4$}\label{higheralg}

As we have explained earlier, the correction functions $F^\rho_{ij}$, that turn the fields $\mathscr T^{E_{ij}}$ into symmetries of the Toda field can be constructed for any ``lowest weight'' representation $\rho$ of the algebra \slnr. In this appendix we study the fields, associated with the $\rho=\rho^6$, the rank $6$ representation of $\mathfrak{sl}_4(\R)$ in $\bigwedge^{2}\R^4$. Let $v_1,\,v_2,\,v_3,\,v_4$ be the standard basis in $\R^4$. Then we have the following basis $f_1,\dots,f_6$ of the representation space $\bigwedge^{2}\R^4$:
\[
\begin{aligned}
f_{1}&=v_{1} \wedge v_{2}, & f_{2}&=v_{1} \wedge v_{3}, & f_{3}&=v_{1} \wedge v_{4},\\
f_{4}&=v_{2} \wedge v_{3}, & f_{5}&=v_{2} \wedge v_{4}, & f_{6}&=v_{3} \wedge v_{4},
\end{aligned}
\]
We will usually denote these vectors by $f_{kl}=v_k\wedge v_l$. The matrix coefficients of this representation are given by the $2\times 2$ minors of the matrices and the lowest weight vector in $\bigwedge^{2}\R^4$ is $f_{34}=v_3\wedge v_4$. Then
\beq{Function-n}
F_{kl} = \langle \rho^{6}(\Psi) f_{kl}, f_{34}\rangle = M_{kl}.
\eeq
Here we denote the minors of the matrix $\Psi\in SO_4(\R)$, spanned by the last two rows and the $k$-th and $l$-th columns of $\Psi$ by
\beq{Minor22}
M_{\frac{34}{kl}} \equiv M_{kl}.
\eeq
As we see, these minors are functions $F$ analogous to the elements $\psi_{ni}$ in the case of tautological representation. On the other hand, we have the following relations
\[
E_{ij}^T v_{k}= E_{ji} v_{k} = \delta_{ik} v_{j},
\]
where $E_{ij}$ is the matrix unit (the matrix with $1$ in the intersection of the $i$-th row and the $j$-th column) in tautological representation of $\mathfrak{sl}_{4}(\mathbb{R})$. Hence in $\bigwedge^2\R^4$ we have
\[
\rho^{6}(\theta (E_{ij})) \circ (v_{k} \wedge v_{l}) = \delta_{ik} v_{j} \wedge v_{l} + \delta_{il} v_{k} \wedge v_{j}.
\]
It follows (see \cite{CSS23}) that the action of vector field $\tc^{E_{ij}}$ on $M_{kl}$ takes the following form:
\beq{Field-action-rho6-1}
\begin{split}
\tc^{E_{ij}} (M_{kl})&=g_{\rho^{6},E_{ij}}(\psi) M_{kl} - \langle \rho^{6}(\Psi)(\rho^{6}(\theta (E_{ij}))(f_{kl})), f_{34}\rangle =\\
&= (\psi_{3i}\psi_{3j}+\psi_{4i}\psi_{4j})M_{kl} - (\delta_{ik}M_{jl} + \delta_{il}M_{kj})
\end{split}
\eeq
and so
\[
\tc^{E_{ij}}\left( \frac{M_{ka}}{M_{la}}\right)= -\delta_{ik}\frac{M_{ja}}{M_{la}}
 -\delta_{ia} \frac{M_{kj}}{M_{la}} + \delta_{il} \frac{M_{ja}}{M_{la}}\frac{M_{ka}}{M_{la}} + \delta_{ia} \frac{M_{lj}}{M_{la}}\frac{M_{ka}}{M_{la}}.
\]
Finally for the commutators we get the following identity
\beq{Commutator-1}
\begin{aligned}
\Biggl[ \frac{M_{ia}}{M_{ja}}\tc^{E_{ij}}, \frac{M_{kb}}{M_{lb}}&\tc^{E_{kl}} \Biggr]=\\
&= \frac{M_{ia}}{M_{ja}}\frac{M_{kb}}{M_{lb}} \cdot
\Biggl(
\left(\delta_{il} \frac{M_{jb}}{M_{lb}} + \delta_{ib} \frac{M_{lj}}{M_{lb}} -\delta_{ik}\frac{M_{jb}}{M_{kb}}
 -\delta_{ib} \frac{M_{kj}}{M_{kb}} \right) \tc^{E_{kl}}\\
 &\quad- \left(\delta_{kj} \frac{M_{la}}{M_{ja}} + \delta_{ka} \frac{M_{jl}}{M_{ja}} -\delta_{ki}\frac{M_{la}}{M_{ia}}
 -\delta_{ka} \frac{M_{il}}{M_{ia}} \right)\tc^{E_{ij}} \Biggr)\\
&\quad+ \frac{M_{ia}}{M_{ja}}\frac{M_{kb}}{M_{lb}} \cdot \left( \delta_{kj} \tc^{E_{il}} - \delta_{il} \tc^{E_{kj}}
 \right).
 \end{aligned}
\eeq
Put
\[
T^{ij}_{a} \equiv \frac{M_{ia}}{M_{ja}}\tc^{E_{ij}}, \ \  T^{kl}_{b} \equiv \frac{M_{kb}}{M_{lb}}\tc^{E_{kl}}
\]
Using the Plukker's identity $M_{12}M_{34} - M_{13}M_{24} + M_{14}M_{23}=0$, we get the Poisson brackets between $T^{ij}_{a}$ and $T^{ij}_{b}$, for example:
\[
\begin{aligned}
\left[ T^{12}_{3}, T^{12}_{4} \right]&= T^{12}_{4} - T^{12}_{3}, &
\left[ T^{34}_{2}, T^{34}_{1} \right]&= T^{34}_{1} - T^{34}_{2}, &
\left[ T^{23}_{4}, T^{23}_{1} \right]&=  T^{23}_{1} - T^{23}_{4}, \\
\left[ T^{13}_{2}, T^{13}_{4} \right]&= -T^{13}_{2} - T^{13}_{4}, &
\left[ T^{12}_{3}, T^{23}_{4} \right]&= T^{13}_{2} + T^{13}_{4}, &
\left[ T^{12}_{4}, T^{23}_{4} \right]&= T^{13}_{4} - T^{12}_{4}, \\
\left[ T^{12}_{3}, T^{23}_{1} \right]&= T^{23}_{1} + T^{13}_{2}, &
\left[ T^{12}_{3}, T^{34}_{2} \right]&= T^{12}_{4} - T^{12}_{3}, &
\left[ T^{12}_{4}, T^{34}_{1} \right]&= T^{34}_{1} - T^{34}_{2},\\
\left[ T^{23}_{4}, T^{34}_{1} \right]&= T^{24}_{1} + T^{24}_{3}, &
\left[ T^{23}_{4}, T^{34}_{2} \right]&= T^{24}_{3} + T^{34}_{2}, &
\left[ T^{23}_{1}, T^{34}_{1} \right]&= T^{23}_{1} + T^{24}_{1},\\
\left[ T^{12}_{3}, T^{13}_{4} \right]&= -T^{13}_{2} - T^{13}_{4}, &
\left[ T^{12}_{4}, T^{13}_{2} \right]&= T^{12}_{3} - T^{12}_{4}, &
\left[ T^{12}_{4}, T^{13}_{4} \right]&= T^{12}_{4} - T^{13}_{4},
\end{aligned}
\]
and further
\[
\begin{aligned}
\left[ T^{12}_{4}, T^{23}_{1} \right]&= \frac{M_{14}}{M_{13}}\frac{M_{23}}{M_{24}}T^{23}_{1} - \frac{M_{12}}{M_{13}}\frac{M_{34}}{M_{24}}T^{12}_{4} +  \frac{M_{12}}{M_{13}}\frac{M_{34}}{M_{24}}T^{13}_{4},\\
\left[ T^{23}_{1}, T^{34}_{2} \right]&= \frac{M_{12}}{M_{13}}\frac{M_{34}}{M_{24}}T^{34}_{2} + \frac{M_{14}}{M_{13}}\frac{M_{23}}{M_{24}}T^{23}_{1} + \frac{M_{14}}{M_{13}}\frac{M_{23}}{M_{24}}T^{24}_{1}
\end{aligned}
\]

\[
\left[ T^{12}_{4}, T^{34}_{2} \right]= 0, \ \ \left[ T^{12}_{3}, T^{34}_{1} \right]=  \frac{M_{12}}{M_{14}}\frac{M_{34}}{M_{23}}(T^{34}_{1} - T^{12}_{3}), \ \
\left[ T^{12}_{3}, T^{13}_{2} \right]= 0.
\]

We see that some of the Poisson brackets depend on the functions:
\[
\frac{M_{14}}{M_{13}}\frac{M_{23}}{M_{24}}, \ \ \frac{M_{12}}{M_{13}}\frac{M_{34}}{M_{24}}.
\]
These functions are invariants under the Toda flow. Thus the invariants of the flow show up in the study of the symmetries of a system, which is to be expected.

\section*{Acknowledgements}
The work of D.T. was carried out within the framework of a development programme for the Regional Scientific and Educational Mathematical Center of Yaroslavl State University, with financial support from the Ministry of Science and Higher Education of the Russian Federation (Agreement on provision of subsidy from the federal budget No. 075-02-2025-1636). The work of G.S. was partly supported by the RSF grant 25-11-00210.

\end{document}